# Fast-Neutron Scintillation with Multi-Quantum-Well 2D Perovskites

*Shaun Tan,[1,†] Zachary Benesh,[1,†] Seong Eui Chang,[1,2] Shelby Elder,[1] Haidi Jakovic,[1,3] Tae-Woo Lee,[2,4,5] Moungi G. Bawendi[1,*]*

[1]Department of Chemistry, Massachusetts Institute of Technology, Cambridge, MA 02139, USA

[2]Department of Materials Science and Engineering, Seoul National University, Gwanak-gu, Seoul 08826, Republic of Korea

[3]Dipartimento di Scienza dei Materiali, Università degli Studi Milano - Bicocca, Milano 20125, Italy

[4]SN Display Co. Ltd., Building 35, Gwanak-gu, Seoul 08826, Republic of Korea

[5]Institute of Engineering Research, Research Institute of Advanced Materials, Soft Foundry, Seoul National University; Seoul 08826, Republic of Korea

†These authors contributed equally to this work.

*Email: mgb@mit.edu

**ABSTRACT**

Fast-neutron scintillators must efficiently convert neutron energy into detectable light while enabling selectivity between neutrons and the $\gamma$-ray background. Here, we report a synthesis strategy for layered two-dimensional (2D) lead halide perovskite scintillators with independent control of the macroscopic detector geometry and microscopic phase composition. Our synthesis approach produces mm-thick, large-area scintillators with a deliberately engineered multi-quantum-well energy landscape, creating an effective spectral separation while retaining nanosecond emission dynamics. Our scintillators exhibit an energy resolution of 5.4% at 662 keV under $^{137}$Cs irradiation. Under deuterium-deuterium fusion neutron irradiation, we directly demonstrate neutron and $\gamma$-ray event separation by pulse shape discrimination, with a figure of merit of 3.51. Taken together, these results establish multi-quantum-well 2D perovskites as a scalable materials platform that combines fast-neutron sensitivity and selectivity within a controllable detector form factor.

**INTRODUCTION**

MeV-scale fast-neutron detection is essential for nuclear energy systems.[1–6] In fusion energy, measurements of neutron yield and energy provide direct information on reaction conditions, such as the deuterium-tritium fuel ratio, ion temperature, and fusion power output.[2,5–7] In the surrounding reactor environment, neutron sensing is necessary for radiation monitoring and operational safety for personnel. Beyond nuclear energy, neutron sensing supports nuclear materials identification, radiation therapy, particle physics, and neutron-based materials characterization.[8–10] These applications motivate scintillators that effectively detect fast-neutrons while operating in mixed radiation fields.

Two-dimensional (2D) lead halide perovskites provide both fast-neutron and $\gamma$-ray sensitivity within a single material platform.[11,12] The organic sublattice enables neutron energy deposition through elastic scattering, while the inorganic lead halide framework converts deposited energy into scintillation light. However, the high-$Z$ inorganic framework also strongly interacts with $\gamma$-rays, making neutron/$\gamma$-ray ($\boldsymbol{n/\gamma}$) particle discrimination an important requirement for sensing in mixed radiation fields. Two coupled challenges thus emerge: (1) Achieving a sufficient neutron interaction volume, and (2) Efficiently outcoupling scintillation photons through the interaction volume.[13–15]

Fast-neutron interaction probability depends strongly on the hydrogen number density and path length through the scintillator.[16] Increasing scintillator thickness enhances neutron interaction, while larger lateral dimensions increase the active detection area. However, conventional 2D perovskite crystal growth preferentially forms thin plate-like morphologies because of anisotropic in-plane crystallization, limiting independent control over scintillator thickness and lateral geometry. Increasing thickness introduces the second challenge associated with optical absorption. Because halide perovskites commonly exhibit relatively small intrinsic Stokes shifts, as the optical path length increases, repeated reabsorption reduces the amount of light collection. Thus, simply increasing thickness to improve neutron interaction will not necessarily improve scintillation detection. Overcoming this geometry-optical outcoupling tradeoff requires independent control over both the scintillator form factor and the excited state energy landscape.

Here, we report a synthesis strategy that independently controls the microscopic phase composition and macroscopic detector geometry of layered 2D perovskite scintillators. Deliberate phase design produces a multi-quantum-well structure where narrower bandgap domains serve as dominant emissive sites within a wide bandgap matrix. This energy landscape creates an effective spectral separation while preserving rapid emission dynamics. Our millimeter-thick scintillators exhibit an energy resolution of 5.4% at 662 keV under $^{137}$Cs irradiation. More importantly, direct measurements under deuterium-deuterium fusion neutron irradiation demonstrate $\boldsymbol{n/\gamma}$ pulse shape discrimination with a figure of merit of 3.51. Combined with analytical calculations and Monte Carlo simulations of neutron interaction and energy deposition, we establish design principles for 2D perovskite scintillators operating in mixed radiation fields.

## RESULTS AND DISCUSSION

Our materials design targets simultaneous control of neutron interaction volume and the optical properties of the scintillator. We thus designed a 2D perovskite architecture where the macroscopic geometry is varied independently of the microscopic phase composition. A predominantly wide bandgap matrix containing a small population of lower bandgap domains provide spectral separation of emission and absorption. This decouples detector form factor, which determines neutron interaction and energy deposition, from phase composition, which sets the excited state energy landscape and photon outcoupling. We first established a synthesis and fabrication process capable of implementing this design over millimeter-scale thicknesses and centimeter-scale lateral dimensions (Figure 1a).

The 2D perovskite precursor solution in *N*,*N*-dimethylformamide is uniformly coated onto a substrate or cast within a mold. Vacuum drying induces solvent removal and 2D perovskite crystallization, and the subsequent thermal annealing further facilitates residual solvent evaporation and crystal growth. Each additional layer is deposited using a precursor solution that is supersaturated at ambient temperature, which kinetically limits redissolution through rapid nucleation and solvent removal, thereby enabling sequential multilayer deposition (Figure S1). This cycle is repeated until the desired scintillator thickness is obtained.

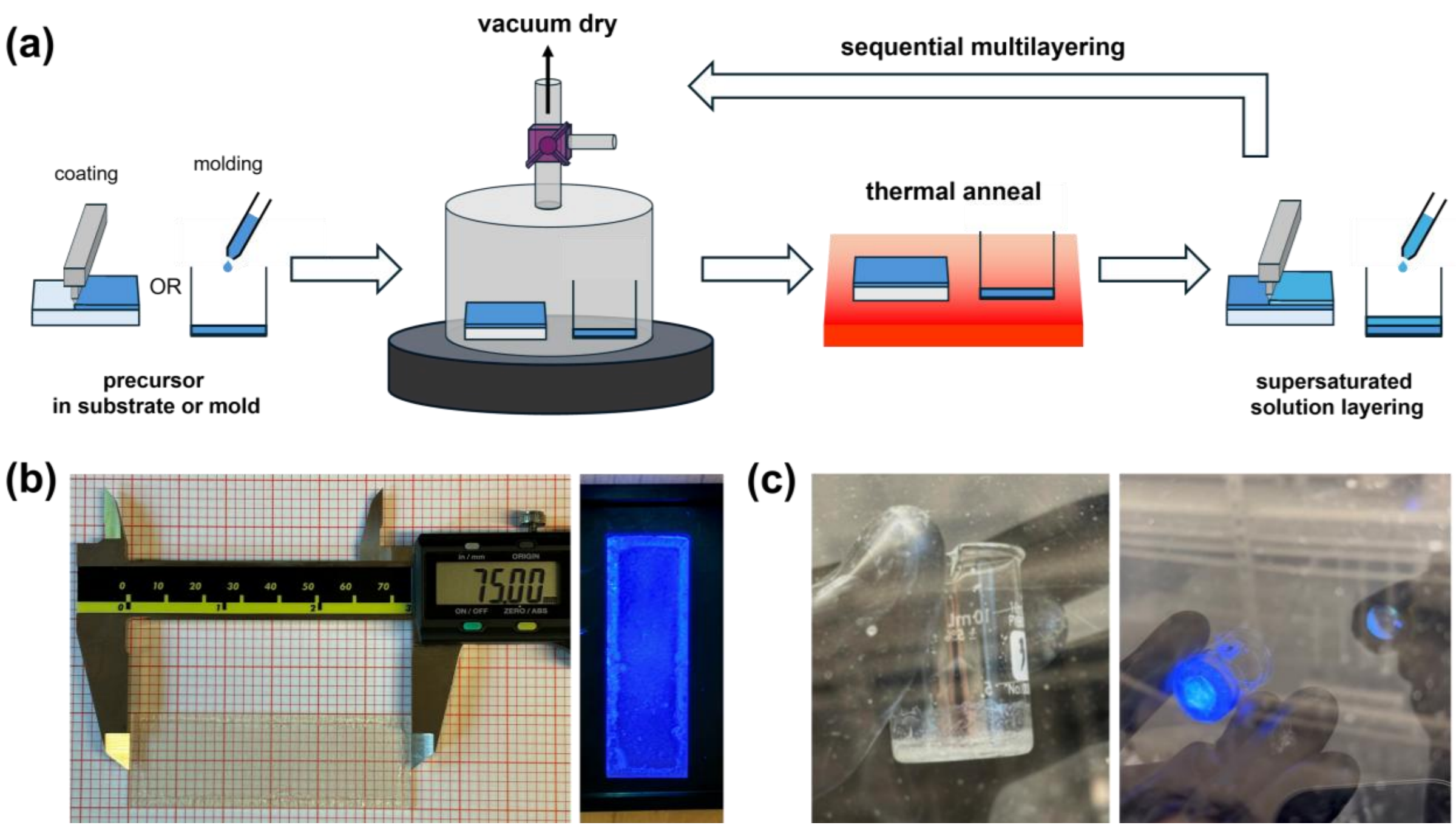

**Figure 1.** (a) Schematic depicting the synthesis approach. Photographs of the (b) rectangular sample prepared on a microscope slide, and the (c) cylindrical sample prepared using a beaker as a mold. Both samples show bright blue photoluminescence under 365 nm ultraviolet excitation.

The macroscopic scintillator geometry is controlled through the substrate area, mold dimensions, and number of deposition cycles. As representative demonstrations, we show a rectangular sample measuring 75×25×0.4 $mm^3$ (length×width×thickness) prepared on a microscope slide (Figure 1b), and a cylindrical disk measuring 25 mm diameter (i.e. 490 $mm^2$ area) and 3.5 mm thickness prepared using a beaker as a mold (Figure 1c). Both samples exhibit bright blue photoluminescence under 365 nm ultraviolet excitation. In particular for the 3.5 mm thick sample, ultraviolet excitation from the back side of the sample produces bright blue emission from the opposite face, qualitatively demonstrating optical outcoupling of the emitted photons across the sample thickness. In comparison, previously reported (Table S1) perovskite scintillators are commonly restricted to sub-millimeter thicknesses.

Our synthesis accommodates off-stoichiometric precursor formulations that do not correspond to a single-phase composition. This contrasts with conventional single crystal growth methods, where the final 2D perovskite is often limited to the stoichiometry of crystallography defined phases and has narrow compositional tolerance, because excess precursor components are excluded from the growing single crystal lattice or segregate into secondary phases. The key advantage of our synthesis is to consolidate multiple compositionally distinct constituents into a continuous material. We demonstrate this compositional flexibility on the archetypal 2D perovskite, $PEA_2PbBr_4$ (PEA = $C_6H_5CH_2CH_2NH_3^+$, phenethylammonium). We explored a range of $PEA_2PbBr_4$-based formulations, containing excess $PbBr_2$, several $APbX_3$ perovskite additives, and different incorporation amounts of $FAPbBr_3$ (FA = $HC(NH_2)_2^+$, formamidinium) (Figure S2). Unless otherwise stated, the discussions henceforth focus on the target formulation of $PEA_2PbBr_4$ with 3 mol% $FAPbBr_3$ and 5 mol% excess $PbBr_2$.

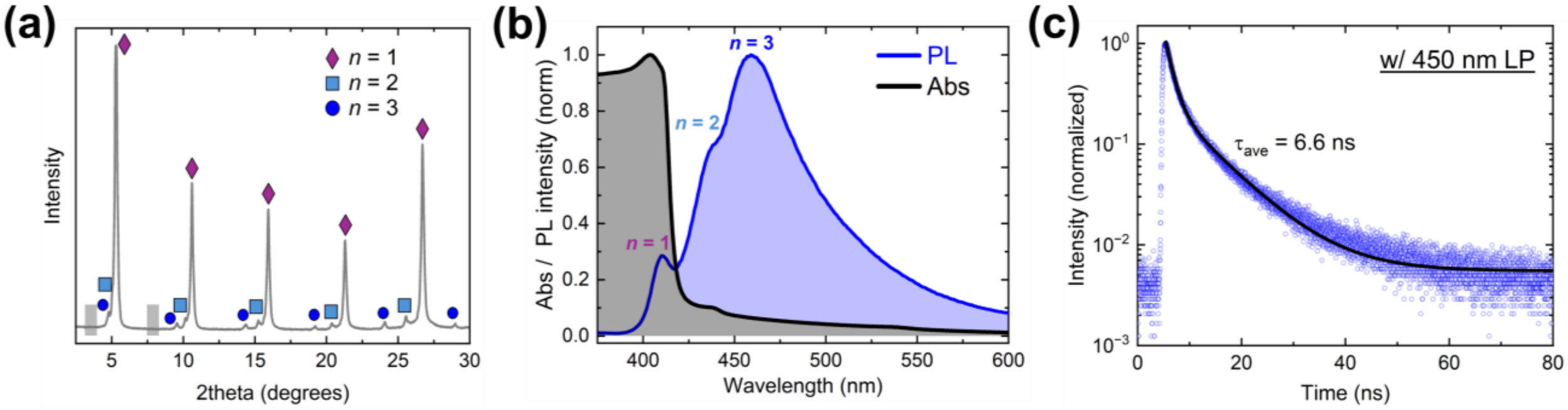

**Figure 2.** (a) XRD pattern of the MQW 2D perovskite. The gray areas represent the (020) and (040) peak positions expected for a stoichiometric 2D(*n*=2) phase. (b) Absorption and photoluminescence spectra of the 2 mm thick MQW sample under 350 nm excitation. (c) TRPL decay of the same sample under 405 nm excitation. The bi-exponential decay fit is denoted by the black line.

The optical bandgap of 2D perovskites generally decreases with increasing inorganic quantum well thickness (i.e. *n*-phase value), because of progressively weaker quantum and dielectric confinement.[17] We thus selected the $FAPbBr_3$ content (i.e. 3 mol% $FAPbBr_3$) to be substantially below the amount required to convert 2D(*n*=1) $PEA_2PbBr_4$ completely into the stoichiometric 2D(*n*=2) $PEA_2FAPb_2Br_7$ phase. The low $FAPbBr_3$ content is thus deliberate to promote localized formation of higher-*n* phases without consuming the majority 2D(*n*=1) component. X-ray diffraction (XRD) of a 2 mm thick sample shows intense periodic Bragg peaks assigned to the 2D(*n*=1) phase (Figure 2a). Meanwhile, the 2D(*n*=2) $PEA_2FAPb_2Br_7$ and 2D(*n*=3) $PEA_2FA_2Pb_3Br_{10}$ phases appear as minority, lower intensity peaks at smaller Bragg angles (i.e. larger d-spacings). Their peak positions are intermediate between those of the 2D(n=1) phase, and what are expected for a stoichiometric 2D(*n*=2) or 2D(*n*=3) phase, suggesting the intergrowth of higher-*n* domains embedded within the 2D(n=1) matrix, rather than the exclusive formation of pure 2D(*n*=2) or 2D(*n*=3) crystallites. Because the relative diffraction intensities are strongly influenced by crystallographic texturing, the XRD pattern is used to identify the coexistence of phases, rather than to quantify their phase fractions.

The absorption spectrum (Figure 2b) provides further evidence that the narrower bandgap higher-*n* domains coexist as the minority component. Despite this, the narrower bandgap domains dominate the photoluminescence spectrum of the same 2 mm thick sample under 350 nm excitation (Figure 2b). The quantum well cascade from the 2D(*n*=1) matrix to the narrower bandgap domains allows for efficient exciton funneling. Therefore, exciton recombination and radiative emission occur preferentially in the narrower bandgap domains. Meanwhile, the lower energy emission also lies largely red shifted from the absorption edge of the 2D(*n*=1) matrix, thereby reducing the spectral overlap by the majority phase. Collectively, the precursor stoichiometry, XRD pattern, and optical spectra support the formation of a wide bandgap 2D(*n*=1) matrix, coexisting with a small fraction of the FA-containing, narrower bandgap, higher-*n* domains. This forms a mixed-phase, multi-quantum well (MQW) 2D perovskite material.

Time-resolved photoluminescence measured using 405 nm excitation, together with a 450 nm long pass filter, shows an average lifetime of 6.6 ns based on a bi-exponential decay fitting (Figure 2c, Table S2). This nanosecond-scale lifetime demonstrates that the spectrally shifted emission does not require a slow, microsecond-scale recombination pathway. While alternative strategies based on transition metal dopants can generate large spectral shifts, their emission lifetimes are on the order of microseconds or longer timescales.[18,19] Overall, our MQW 2D perovskite achieves spectral separation while retaining rapid emission dynamics. This combination is important to preserve the temporal response required for mixed-field $\boldsymbol{n/\gamma}$ pulse shape discrimination,[6,20,21] as demonstrated below under direct fast-neutron irradiation.

Besides the optical properties, the material uniformity also visibly improves through the use of off-stoichiometry compositions (Figure S3). The surface morphology is polycrystalline, with lateral grain sizes exceeding 400 µm (Figure S4). Elemental mapping shows largely homogeneous distribution of the constituent elements, with localized variations along the grain boundaries (Figure S5). We note that elemental mapping does not resolve the spatial distribution of PEA and FA, because the FA concentration is low and both organic cations contain carbon and nitrogen. Scanning probe microscopy (SPM) of a representative grain boundary shows a grain-to-grain step height of approximately 40 nm (Figure S6). Despite the grain boundary features, the overall surface remains smooth, with an average root-mean-square (RMS) roughness of approximately 5 nm over a total measured area of 270 μm$^2$ (Figure S7). This surface uniformity facilitates sequential overcoating to grow thick scintillators.

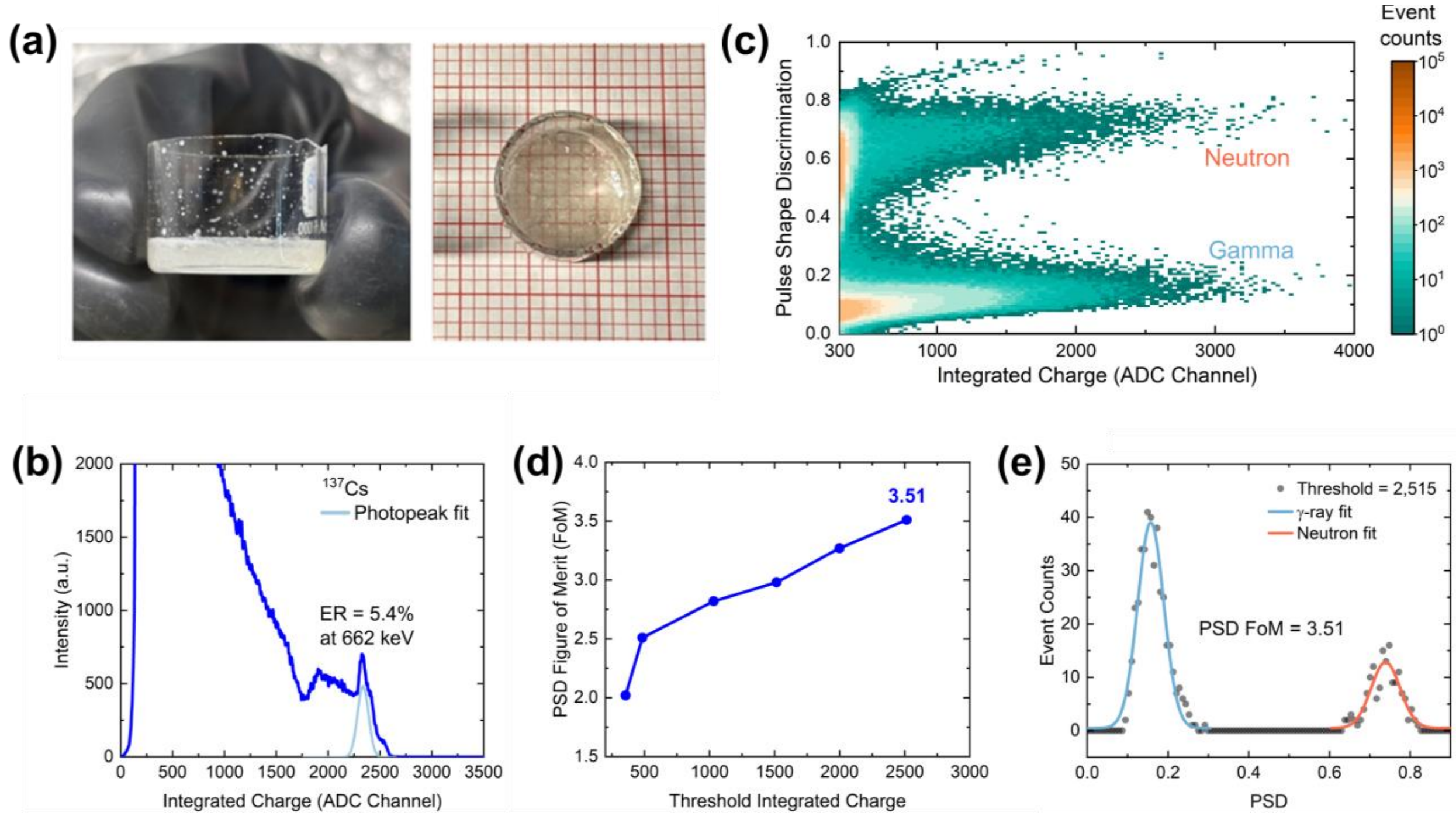

**Figure 3.** (a) Photographs of a representative scintillator used for radiation measurements. (b) Pulse height spectrum under $^{137}$Cs γ-rays. (c) PSD map measured under D-D neutron generator irradiation. (d) $\boldsymbol{n/\gamma}$ discrimination FoM value with increasing threshold integrated charge. (e) $\boldsymbol{n/\gamma}$ PSD histogram at a threshold integrated charge of 2,515. Gaussian fits to the neutron and γ-ray populations are denoted by the solid line fits.

For all scintillation measurements, we used cylindrical samples of 490 mm$^2$ area and 5 mm thickness (Figure 3a). We first evaluate the γ-ray response using a $^{137}$Cs source (Figure 3b). The pulse height spectrum exhibits a well-defined full-energy photopeak at 662 keV with an energy resolution of 5.4%. The pronounced photopeak is enabled by the thick scintillator geometry, which increases both the probability of γ-ray interaction and the likelihood that Pb characteristic x-rays and Compton photons are fully absorbed within the scintillator volume. In contrast, thinner geometries typically show a reduced photopeak convolved with the Pb escape feature, because a larger fraction of γ-rays would escape without depositing their full energy. Relative to an EJ-254 commercial scintillator (Figure S8), our MQW 2D perovskite demonstrates a light yield of 16,868 photons/MeV. The 5 mm thickness was designed to enhance fast-neutron interaction rather than to optimize γ-ray scintillation, as the high-$Z$ inorganic framework provides substantial γ-ray interaction at smaller dimensions. The measured responses thus demonstrate strong γ-ray scintillation and spectral resolution in a detector geometry aimed at fast-neutron sensing.

We evaluate fast-neutron scintillation and $\boldsymbol{n/\gamma}$ discrimination under irradiation from a deuterium-deuterium (D-D) generator, which produces monoenergetic neutrons of 2.45 MeV. During generator operation, the background also consists of γ-rays generated by neutron-induced interactions with the surrounding materials in the environment. This provides a mixed $\boldsymbol{n/\gamma}$ field to evaluate particle discrimination. Pulse shape discrimination (PSD) was quantified using the charge comparison method where,

$$\text{PSD} = \frac{Q_L - Q_S}{Q_L}$$

with short ($Q_S$) and long ($Q_L$) charge integration gates of 22 ns and 160 ns respectively. Neutron events and the γ-ray background are clearly distinguishable by two population bands in the PSD map (Figure 3c). The neutron population extends across a broad range of pulse amplitudes, consistent with a continuum of energy transferred through fast-neutron elastic recoil and partial energy deposition. The neutron population remains concentrated within a distinct PSD range relative to the γ-ray background. We analyze the PSD distributions as a function of pulse amplitude thresholding, with events integrated from the selected threshold to the upper acquisition limit. The figure of merit (FoM) is defined as,

$$\text{FoM} = \frac{|\mu_n - \mu_\gamma|}{FWHM_n + FWHM_\gamma}$$

where $\mu_n$ and $\mu_\gamma$ are the centroids and $FWHM_n$ and F$HMW_\gamma$ are the full-width-at-half-maximum of the neutron and γ-ray Gaussian fits. The FoM compares the separation between the two PSD populations, with a greater centroid separation and narrower distribution thus producing stronger $\boldsymbol{n/\gamma}$ event separation. The FoM increases monotonically with thresholding (Figure 3d, Figure S9), consistent with improved discrimination as low-light scintillation events with larger statistical fluctuations are progressively excluded. At an integrated charge threshold of 2,515, the neutron and γ-ray populations are clearly resolved, reaching a FoM of 3.51 (Figure 3e), demonstrating excellent fast-neutron discrimination from the γ-ray background in our MQW 2D perovskite scintillators.

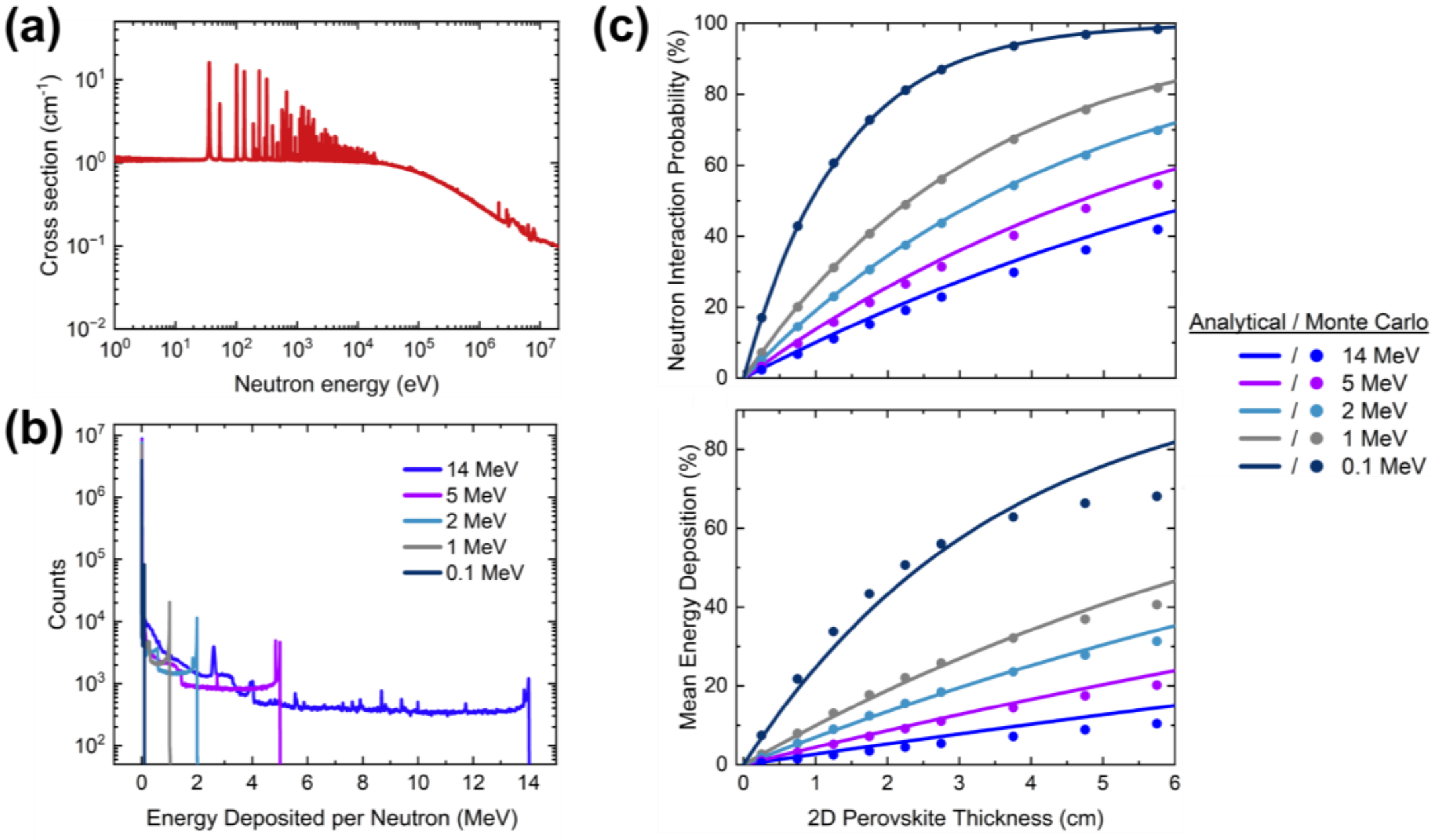

**Figure 4.** (a) Macroscopic total neutron cross-section of $PEA_2PbBr_4$. (b) Monte Carlo simulations of the distribution of energy deposition per incident monoenergetic neutron. (c) Probability of at least one neutron interaction (top panel), and mean energy deposition per incident neutron (bottom panel), as a function of $PEA_2PbBr_4$ thickness.

We quantitatively assess the effect of scintillator thickness on fast-neutron interaction and energy deposition. The macroscopic total neutron cross-section of $PEA_2PbBr_4$ varies strongly with incident neutron energy (Figure 4a). Monte Carlo simulations with monoenergetic neutrons produce broad distributions of deposited energy (Figure 4b). These distributions reflect the stochastic variations in energy transfer per collision, secondary particle transport, and number multi-scattering events, thus yielding a continuum of partial energy deposition events. This is consistent with the broad pulse amplitude distribution we experimentally observed under D-D irradiation (Figure 3c). The probability that an incident neutron undergoes at least one interaction (Figure 4c top panel), and the mean energy deposition (Figure 4c bottom panel), both increase with scintillator thickness. The analytical calculations largely reproduce the Monte Carlo tends, indicating that the first order thickness dependance is largely determined by the energy-dependent total cross section and neutron path length. At 2 MeV, representative of fission neutrons from $^{252}Cf$ or $^{235}U$, and close to the D-D neutron energy (2.45 MeV), our 5 mm thick scintillator has a calculated 10.1% neutron interaction probability, and on average 3.6% of the neutron energy is deposited. This benchmark illustrates the central detector design tradeoff: Increasing thickness enhances neutron interaction and energy deposition but also increases the optical path length and associated reabsorption losses. Our MQW 2D perovskite addresses these competing requirements through independent control of the macroscopic form factor and microscopic phase composition.

More broadly, this work demonstrates that chemically designed 2D perovskites can combine the interaction volume required for fast-neutron sensing, with the optical and temporal response needed for effective neutron pulse shape discrimination from the γ-ray background. Our scintillators directly detect the 2.45 MeV D-D neutrons and achieve a FoM of 3.51 in a mixed radiation field. This capability is particularly relevant for neutron sensing in fusion energy environments, where intense γ-ray backgrounds complicate neutron identification. Beyond the specific performance demonstrated here, our results establish chemical phase engineering and detector form factor control as complementary design strategies for translating halide perovskites from efficient light emitters into particle discriminating radiation detectors, providing a materials design framework for future mixed field radiation sensing.

## SUPPORTING INFORMATION

The Supporting Information is available free of charge at

- Materials and Methods, Sequential multilayering, Off-stoichiometric compositional flexibility, Off-stoichiometric material uniformity, Surface morphology, Elemental spatial distribution, Grain boundary topography, Surface topography, EJ-254 $\gamma$-ray scintillation, Threshold-dependent $\boldsymbol{n}/\boldsymbol{\gamma}$ PSD

## AUTHOR INFORMATION

### Corresponding Authors

**Moungi G. Bawendi** - *Department of Chemistry, Massachusetts Institute of Technology, Cambridge, MA, USA*; Email: mgb@mit.edu

### Authors

**Shaun Tan** - *Department of Chemistry, Massachusetts Institute of Technology, Cambridge, MA, USA*

**Zack Banesh** - *Department of Chemistry, Massachusetts Institute of Technology, Cambridge, MA, USA*

**Seong Eui Chang** - *Department of Chemistry, Massachusetts Institute of Technology, Cambridge, MA, USA, and Department of Materials Science and Engineering, Seoul National University, Gwanak-gu, Seoul 08826, Republic of Korea*

**Shelby Elder** - *Department of Chemistry, Massachusetts Institute of Technology, Cambridge, MA, USA*

**Haidi Jakovic** - *Department of Chemistry, Massachusetts Institute of Technology, Cambridge, MA, USA, and Dipartimento di Scienza dei Materiali, Università degli Studi Milano - Bicocca, Milano 20125, Italy*

**Tae-Woo Lee** - *Department of Materials Science and Engineering, Seoul National University, Gwanak-gu, Seoul 08826, Republic of Korea, and SN Display Co. Ltd., Building 35, Gwanak-gu, Seoul 08826, Republic of Korea, and Institute of Engineering Research, Research Institute of Advanced Materials, Soft Foundry, Seoul National University; Seoul 08826, Republic of Korea*

**Author Contributions**

S.T. and Z.B. contributed equally to this work.

**Notes**

S.T., Z.B., and M.G.B. are inventors of a provisional patent (US 64/142,218) related to the subject matter of this manuscript. The other authors declare that they have no competing interests.

**ACKNOWLEDGMENTS**

S.T., Z.B., and H.J. gratefully acknowledge Eni S.p.A. for funding this research. H.J. was partially supported from The Foundation Blanceflor through grant no. 304-2026:1. S.E.C. was partly supported by a Korea Institute of Energy Technology Evaluation and Planning (KETEP) grant funded by the Korean government (Ministry of Trade, Industry and Energy) (RS-2025-25460428). This work was performed in part in the MIT.nano Characterization Facilities. The authors would like to thank E. Lamere for assisting with the $^{137}$Cs source. The authors would like to thank E. Edwards and M. Johnson for assisting with the D-D neutron generator. The authors would like to thank C. Moorman for assisting with the field-emission SEM measurements.